# The Effect of Client Appraisal on the Efficiency of Micro Finance Bank

Esther Yusuf Enoch[1], Usman Abubakar Arabo[1], Abubakar Mahmud Digil[1]

[1](Department of Banking and Finance, Federal Polytechnic Mubi, Adamawa State, Nigeria)

**ABSTRACT**: *One of the major problems confronting financial institutions most especially microfinance institutions is the increasing incidence of loan defaults and consequence loan losses which manifested in their financial performance with huge uncollectible loans and advances. This study assessed the effects of credit management on financial performance on microfinance institutions in Adamawa State, Nigeria. Specifically, we examine the effect of client appraisal on the efficiency of microfinance banks in Adamawa State. The methodology employed in this study is the survey method in which both primary and secondary sources were used in the collection of data. A multi-stage sampling method was adopted in selecting a sample of 21 respondents from a total population of 52 credit officers. Questionnaires were used in the due collection of data from the respondents. Descriptive statistics (simple percentage) and inferential statistics (regression analysis) were used to analyze the data collected and in testing the hypotheses. The study showed that client appraisal has a positive effect on efficiency and productivity.*

## I. INTRODUCTION

A key requirement for effective credit management is the ability to intelligently and efficiently manage customer credit lines, by having greater knowledge of customer strength, credit score history, and changing payment patterns. Microfinance Institution successes, largely depend on the effectiveness of its credit management, because once these loans are not repayable, the bank faces high credit risk. The biggest risk in microfinance banks is granting credit and not getting it back which causes a total loss to the bank. According to [9], granting credit may accrue benefits of increasing sales to the institution; there are high default risks that may adversely affect its future. That is why microfinance banks need to consider some factors like Client Appraisal where the 5Cs model of credit i.e. character, collateral, and capital, condition, and capacity will be employed to weigh the ability of a potential borrower. Also, debt collection policies and credit risk control measures should be put in place even before embarking on lending, since some customers are slow payers while others are non-payers. These appropriate credit management policies will help in making a profit, reduces the rate of non-payment, and achieve the purpose for which they are established.

In a nutshell, once credit management policy is correctly formulated, carried out, and well understood at all levels of the financial institution, a sound financial performance will be achieved and the complexities involved in planning, executing, and controlling the overall running of the bank will be minimized. Well-founded credit management is a prerequisite for a sustained and profitable financial institution while deteriorating credit quality is a sure cause of poor financial performance and condition. The major cause of serious banking problems continues to be directly related to low credit standards for borrowers, poor portfolio management, and lack of attention to changes in economic or other circumstances that can lead to deterioration in credit quality. Microfinance institutions face the risk of granting credit and not getting it back and in such cases, a total loss is incurred. Also, the risk of delay in repayment (late payment) from debtors as such affects the relationship between the lender and the borrower in which profitability is eroded. The excessively high level of non-performing loans today can also be attributed to poor governance practices, poor credit administration processes, and the absence of non-adherence to credit management policies thereby adversely affecting financial performance.

Fredrick [4] conducted a study on the effect of credit risk management procedures on financial performance among microfinance institutions in Nairobi, the study revealed the various procedures in credit management that will improve financial performance. Wanjira [13] studied the relationship between non-performing loan management practices and the financial performance of commercial banks in Kenya. The study concluded that there is a need for banks to adopt non-performing loan management practices. The study further revealed that there is a relationship between non-performing loan management practices and the financial performance of commercial banks in Kenya. From the above studies, it shows that no study has used portfolio quality (total fund available for MFIs to use as loans in order to satisfy client) and efficiency (time) as indices for measuring the financial performance of microfinance institutions. Most of the studies above showed the procedures, practices, and strategies used by various financial institutions ([4], [13]). The study did not establish a clear relationship between credit management and financial performance. Also, it shows that there is no evidence of empirical studies on the effect of credit management on the financial performance of microfinance institutions in Adamawa state. Hence, this study addresses that gap. Therefore this study assesses the effect of credit management on the financial performance of microfinance institutions in Adamawa state.

The major contribution of this paper is to assess the effect of credit management on the financial performance of microfinance institutions in Adamawa State. In particular, we assess the effect of client appraisal on the efficiency of microfinance banks in Adamawa State, Nigeria.

## II. RELATED WORK

This section discusses the related work on client appraisal, a microfinance bank, and the process of credit management.

Micro Finance Institutions use the five C's of credit to focus on the key dimension of an applicant's creditworthiness. According to [1], Micro Finance Institutions use the 5 Cs model of credit to evaluate a customer as a potential borrower. The 5Cs help MFIs to increase loan performance, as they get to know their customers better. These 5Cs are; Character, capacity, collateral, capital, and condition. A character is a tool that provides weighting values for various characteristics of a credit applicant and the total weighted score of the applicant issued to estimate his creditworthiness [1]. It is focused on the past payment history and the integrity of the potential borrower.

Factors that influence a client can be personal or social factors [11]. The psychological factor basically is the client's inner worth rather than his tangible evidence of achievements. Micro Finance under this factor considers the potential borrower's attitude and his past repayment of the loan granted to him. Under personal factors, it includes age, life cycle stage, occupation, etc. the Social factors involve membership groups, consumption, and entertainment.

Capacity: This is the applicant's ability to repay the requested credit. The MFI's will consider the cash flow from the business, the timing of repayment, and repayment of the loan without any default. Anthony [2] defines cash flow as the cash a borrower has to pay his debt. The cash flow assists the Micro Finance Institution to ascertain if the potential borrower will be able to pay the loan granted to him/her without breaching. Financial statement analysis, with particular emphasis on liquidity and debt ratios, is used to asset the borrower's capacity.

Collateral: This is an asset that a customer has to pledge against debt. Collateral also, is the assets that the potential borrower has in his possession in securing credit. These assets can be used to repay the loan in case the borrower defaults on his payment.

Capital: This is the financial strength of the potential borrower (i.e., assets and liabilities of the business). Haron, *et al.* [7] define capital as the money a borrower has personally invested in the business and is an indication of how much the borrower has at risk should the business fail.

Condition: Condition refers to the borrower's sensitivity to external forces such as interest rates, inflation rates, business cycles, and competitive pressures. The current economic and business climate as well as any unique circumstances affecting either party to the credit transaction are what to be considered before granting credit to a customer/client. Micro Finance Institutions (MFIs) give more attention to character and capacity because the two C's of credit are the most basic requirement to be considered before advancing any loan to a potential borrower. The other 3C's are important in structuring the credit management and making the final decision.

A. Client Appraisal and Efficiency of Microfinance Banks

A key necessity for viable credit risk management is the ability to efficiently oversee the client credit line. To minimize the introduction of terrible obligations, over-saving, and liquidation, a bank must have a more prominent understanding of client budgetary quality, financial assessment history, and changing installment designs[10]. Credit management for a loan deal does not stop until the last or full installment has been recovered. For banks to ensure efficiency it must have greater insight into important factors like customers financial strength, credit score history, and changing payment pattern

Loan appraisal is an application/request for funds, evaluated by financial institutions. The aspect to be focused in appraisal includes; the borrower, the business of the borrower, the capital resources of the borrower, the amount required, the purpose of the loan, the sources of repayment, the security, and the profitability. Loan appraisal plays an important role to keep the loan losses to a minimum level, hence if the officer appointed for loan appraisal is not competent then there will be high chances of lending money to non-deserving customers [3].

Credit risk is a risk of borrower default, which happens when the counterpart fails to pay on time. There can be many reasons for default. One of the most common ones is the obligor in a financially stressed situation [6], besides, if a borrower with a high credit quality has a deteriorated profile, it can also lead to credit risk loss to the bank. If disbursed loans are successfully recovered by the granting financial institution then efficiency is been achieved, also good and competent loan officers play a prominent role in enhancing the recovery of loans while obtaining reasonable collateral from the client will also help banks to avoid losses.

Cultivating good loan customers and using credit risk analysis to ensure that borrowers are creditworthy is central to banks long term profitability. Client appraisal could lead to efficiency in MFIs since it directly affects the progress of microfinance institutions in terms of both officers and other personnel of the institutions. Where clients are properly assessed, the tendency to contract losses caused by nonpayment of loans is less, consequently, efficiency in banks will be high.

B. Efficiency

The efficiency of Micro Finance Institutions is measured by the share of operating expense to the gross loan portfolio. The ratio provides a broad measure of efficiency as it assesses both administrative and personnel expense with lower values indicating more efficient operation.

$$\text{Operating expenses} = \frac{\text{Personnel and administrative expense}}{\text{Period} - \text{Average gross loan portfolio}}$$

Efficiency indicators reflect how well Micro Finance Institutions use their resources, particularly their assets and personnel.

It helps institutions to have a competitive edge in the industry. Management decisions about credit methodology, credit terms, and the market in which to operate directly affect efficiency.

The operating expense ratio enables managers to compare quickly administrative and personnel expenses to the MFIs yield on the gross loan portfolio. Therefore the operating expenses ratio is frequently referred to as the efficiency ratio monitoring this trend is an easy way to observe if the MFI is increasing its efficiency as it grows its loan portfolio.

The lower the ratio, the more efficient the MFIs are. That is why is paramount for MFIs to strive for a downward trend in this ratio –even when portfolio growth is flat until they are convinced that no more efficiencies can be found.

The debt-equity ratio is a member of the asset/liability management ratios and specifically attempts to track MFI's leverage. These measures provide information on the capital adequacy of MFIs and assess the susceptibility to the crisis. Micro Finance mainly relies on this ratio because it assists in predicting the probability of Micro Finance honoring its debt obligation.

The operating expenses ratio is the most widely used indicator of efficiency, but its substantial drawback is that it will make an MFI small loans look worse than an MFI making large loans, even if both are efficiently managed. Thus, a preferable alternative is a ratio that is based on clients served, not the amount loaned.

$$\text{Cost per client} = \frac{\text{Operating expenses}}{\text{The average number of active clients}}$$

The cost per client ratio indicates to an institution how much it currently spends on personnel and administrative expenses to serve a single active client. It tells MFI how much it must earn from each client to be profitable.

### III. METHODOLOGY

The methodology employed in this study is the survey method in which both primary and secondary sources were used in the collection of data. A multi-stage sampling method was adopted in selecting a sample of 21 respondents from a total population of 52 credit officers. Questionnaires were used in the due collection of data from the respondents. Descriptive statistics (simple percentage) and inferential statistics (regression analysis) were used to analyze the data collected and in testing the hypotheses. We used the software SPSS (Statistical Package for Social Science) to implement the statistical techniques mentioned above.

The hypothesis for this research is - Client Appraisal has no significant effect on the Efficiency of Micro-Finance banks. This hypothesis will be tested using multiple regressions. In the following, we describe the sources of data, population, and method of testing the hypothesis.

A. Sources of Data

The sources of data are employees of microfinance institutions in Adamawa state, in which the primary and secondary data were obtained. The primary data consist of the questionnaires which were distributed to the respondents and secondary data comprised of textbooks, downloaded internet materials as well as annual reports and bank sources (loan application forms and loan appraisal forms) were all used to achieve the research result.

B. The population of the Study

In Adamawa State, there are eight (8) licensed Micro Finance Banks as of the time for this research. This research assesses the effect of Credit Management on the financial performance of Micro Finance Banks in Adamawa state for a period of four years (2010 – 2016), the period was selected due to substantial growth in small-scale businesses in Adamawa State which increased the demand for credit facilities. These banks include;

Biyama Micro Finance Bank Limited, Hong LGA, Bonghe Micro Finance Bank Limited, Numan LGA, Fufore Micro-Finance Bank Limited, Fufore LGA, Girei microfinance bank Girei LGA, Gudusisa Micro Finance Bank Limited, Gombi LGA, Michika Micro-Finance Bank, Michika LGA, Standard Micro-Finance Bank Limited, Yola North LGA and Ummah Micro Finance Bank, Yola South LGA which covers the northern, central and southern senatorial district of Adamawa State. However, the only MFI in the Northern zone (Michika Micro Finance Bank) could not be accessed due to insecurity. The researcher only focused on the seven banks located in the central and southern senatorial zone.

The population of the study is the eight (8) licensed Micro Finance institutions, which comprised of 51 employees (credit officers) of all the eight microfinance institutions in Adamawa State, Nigeria.

Sample Size and Sampling Technique: The sample size of the study is the total population size. For the selection of banks, the Purposive Sampling Technique was employed and this was a result of the small population size.

The sampling frame for the research was structured in such a way that questionnaires was administered to Credit Officers of the eight licensed microfinance banks in Adamawa State as reflected below:

Table 1. Schedule of Population Frame

| S/No | Senatorial District | LGA | Name of Bank | No. of Credit Officers |
|---|---|---|---|---|
| 1. | Southern District | Numan | Bonghe Micro Finance Bank Ltd | 8 |
| 2. | Central District | Yola North | Standard Micro Finance | 8 |
| | | | Ummah Micro Finance Bank Ltd | 8 |
| | | | Girei Micro Finance Bank Ltd | 7 |
| | | | Fufore Micro Finance Bank Ltd | 7 |
| | | | Gudusisa Micro Finance Bank Ltd | 8 |
| | | | Biyama Micro Finance Banke Ltd | 5 |
| 3. | Northern District | Michika | Michika Community Bank Ltd Not accessible | Nil |
| Total Number of Respondents | | | | 51 |

**Data Source: Survey Work 2020**

  C. Method of Data Collection

The method of data collection was the survey method. Structured questionnaires were distributed to credit officers of the selected microfinance banks in order to assess the effect of credit management on the financial performance of microfinance institutions. The questionnaire was divided into two sections. Section A was concerned with the demographic and general information about respondents; section B addressed the study objectives were designed in such a way that all the required information obtained from the administration of structural closed-ended questionnaires.

  D. Method of Data Analysis

Data were analyzed using Descriptive and Multiple Regression. The analyses were done through the aid of Statistical Package for Social Science (SPSS) Version 17. Multiple Regression analyzes the hypothesis. The model was used in an attempt to assess the effect of Credit Management on the financial performance of Microfinance Institution in Adamawa State. The regression used is given below;

$$Y = a + B_1X_1 + B_2X_2 + B_3X_3 + e_i$$

Where;
Y= Dependent variable
a=Constant (Intercept)
$B_i$= Coefficient of independent variables (i=1, 2, 3,4...)
$X_i$=Independent variables (i=1, 2, 3,4, 5...)
$X_1$= Client Appraisal
$X_2$= Collection Policy
$X_3$= Credit Risk Control
$e_i$= error term

# IV. DATA PRESENTATION AND ANALYSIS

This section presents and analyses the data collected through questionnaires, and the formulated hypotheses were tested and interpreted. A total number of fifty-one (51) questionnaires were distributed to the credit officers of seven selected microfinance banks in Adamawa State. They were filed and returned because of adequate time given to them. . However, prior to the test of each hypothesis, Pearson's correlation of both one and two-tailed significance was employed to explain the association between the regresses and the regressors as well as the association between the regressors themselves. Similarly, the Kolmogorov – Smirnov normality test was used to test whether the sample distribution is significantly different from a normal distribution. Finally, multiple regression was used to test the hypothesis.

Data Presentation, Analysis, and Interpretation

An empirical analysis and interpretation were done from the data collected through the designed questionnaires. They are presented serially in tables in accordance with each question as below.

Table 2. Distribution of the Respondents according to Sex

| Variable | Frequency | Percentage |
|---|---|---|
| Male | 37 | 72.5 |
| Female | 14 | 27.5 |
| Total | 51 | 100.0 |

*Data Source: Survey Work 2020.*

Table 2 indicates the distribution of the respondents according to their sex. Respondents that are male constitute the majority with about 37 representing 72.5%. However female respondents accounted for the least with 14 denoting 27.5%. This shows that the majority of the respondents are male. This might not escape from the fact that most of our workforce are males. The fact that they are male more production is expected as men are less busy with household responsibility and maternity leaves and so on.

Table 3. Distribution of Respondents base on Age

| Variable | Frequency | Percentage |
|---|---|---|
| 18 – 25 years | 11 | 21.6 |
| 26 – 33 years | 21 | 41.2 |
| 34 – 41 years | 10 | 19.6 |
| 42 – 49 years | 5 | 9.8 |
| 50 + years | 4 | 7.8 |
| Total | 51 | 100.0 |

*Data Source: Survey Work 2020.*

Table 3 shows the distribution of respondents according to age. Those who are in the age range of 26 – 33 years formed the majority of the respondents with 21 portraying 41.2%. This is seconded by those between the ages of 18 – 25 years with 11 representing 21.6%. Those within the age bracket of 34 – 41 years accounted for the third position with 10 representing 19.6%. However, those within the age bracket of 42 – 49 and 50 years above constituted the least with 5 and 4 respondents connoting 9.8% and 7.8%. This implies that most of the credit management administrators of the banks are in their active age. Therefore full productivity is expected in the respective microfinance banks.

Table 4. Distribution of Respondents according to Name of their Banks

| Variable | Frequency | Percentage |
|---|---|---|
| Biyama | 5 | 9.8 |
| Bonghe | 8 | 15.7 |
| Furore | 8 | 15.7 |
| Girei | 7 | 13.7 |
| Gudusisa | 8 | 15.7 |
| Standard | 8 | 15.7 |
| Ummah | 7 | 13.7 |
| Total | 51 | 100.0 |

*Data Source: Survey Work 2020.*

Table 4 presents the names of banks of the respondents, with those coming from Bonghe, Furore, Gudusisa, and Standard Microfinance banks accounted for the highest with 8 denoting 15.7% each. Respondents coming from Girei and Ummah microfinance banks accounted for the second with 7 connoting 13.7% each. However, opinions coming from respondents from Biyama Microfinance recorded as the least with 5 revealing 23.8%. This means that the banks are proportionally or equally represented in respect of credit management.

Table 5. Distribution of Respondents base on Years of Service

| Variable | Frequency | Percentage |
|---|---|---|
| 1 – 5 years | 28 | 54.9 |
| 6 -10 years | 14 | 27.5 |
| 11 – 15 years | 8 | 15.7 |
| 16+ years | 1 | 2.0 |
| Total | 51 | 100.0 |

*Data Source: Survey Work 2020.*

Table 5 depicts the distribution of respondents according to the number of years served in their respective banks. Respondents whose years of service in bank range from 1 – 5 years reported as the highest with 28 revealing 54.9%. It is followed by those who served in the banks between 6 – 10 years with 14 representing 27.5%. However, those who served in the banks for 11 – 15 considered as the third with 8 depicting 15.7%. Meanwhile, respondents who served for 16 years and above considered as the least with 1 displaying 4.8%. The fact that majority of the respondents have average years of service in their respective banks; they have average experience in credit management thereby making the opinion objectively.

Table 6. The Existence of Microfinance Banks

| Variable | Frequency | Percentage |
|---|---|---|
| Less than 5 years | 0 | 00.0 |
| Between 5 – 10 years | 13 | 25.5 |
| Between 10 – 15 years | 38 | 74.5 |
| Above 15 years | 0 | 00.0 |
| Total | 51 | 100.0 |

*Data Source: Survey Work 2020.*

Table 6 reveals how long the microfinance banks of the respondents were being in existence. Respondents, that confirmed that their bank is in existence between 10 – 15 years accounted for the highest with 38 people representing 74.5%. This is because microfinance banks started operation in Nigeria in the year 2005 and some of them were in existence before as community banks. However, 13 respondents denoting 25.5% attested that they were in the business between 5 – 10 years. This implies that almost all the banks were in existence for over 10 years therefore might have a well-structured credit management policy.

Table 7. Adoption of Credit Management Practice

| Variable | Frequency | Percentage |
|---|---|---|
| Yes | 51 | 100.0 |
| No | 0 | 00.0 |
| Total | 51 | 100.0 |

*Data Source: Survey Work 2020.*

Furthermore, Table 7 shows the adoption of credit management practices by microfinance banks. It indicates that all the banks adopt credit management practice with 51 respondents representing 100%. However, there is no response with regard to variable 'NO' which suggests the absence of credit management practice by the banks. This means that all the banks engage in credit management practice, perhaps the practice may vary from one bank to another.

Table 8. Number of Clients by respective Microfinance Banks

| Variable | Frequency | Percentage |
|---|---|---|
| Less than 100 | 4 | 7.8 |
| Between 100 - 200 | 2 | 3.9 |
| Between 200 - 300 | 1 | 2.0 |
| Above 300 | 44 | 86.3 |
| Total | 51 | 100.0 |

*Data Source: Survey Work 2020.*

Table 8 depicts the number of clients possesses by each microfinance bank in the state. Microfinance banks that possess more than 300 clients accounted for the highest with 44 respondents denoting 86.3%. Meanwhile, those with less than 100 clients have 4 representing 7.8%. However, those with 100 - 200 clients recorded as the second least with 2 revealing 3.9%. Those in between 200 - 300 recorded as the least with 1 each representing 2.0%. This means that majority of the banks have more than 300 clients, therefore credit management practice is essential in order to avoid default payment. Equally, it helps to maintain liquidity in the banks.

Table 9. Efficiency and Productivity of Banks Credit Management before Clients Appraisal

| Variable | Frequency | Percentage |
|---|---|---|
| Poor | 17 | 33.3 |
| Moderate | 25 | 49.1 |
| Strong | 9 | 17.6 |
| Total | 51 | 100.0 |

*Data Source: Survey Work 2020*

Table 9 above shows the rate of efficiency of microfinance banks' credit management before clients' appraisal. Those that viewed the credit management policies of the banks as moderate recorded as the highest with 25 people connoting 49.1%. This was seconded by respondents who considered it as poor with 17 representing 33.3%. However, those that rated it as strong formed the least score with 9 denoting 17.6%. This implies that prior to clients' appraisal the credit management policy of the banks is moderate. So there is a need for client appraisal so as to have an efficient credit management policy.

Table 10. Rate of Credit Management under Character Appraisal of Clients

| Variable | Frequency | Percentage |
|---|---|---|
| Poor | 5 | 9.8 |
| Moderate | 10 | 19.6 |
| Strong | 36 | 70.6 |
| Total | 51 | 100.0 |

*Data Source: Survey Work 2020*

Table 10 shed more light on how respondents rated the credit management policy of their respective banks under the character appraisal of the client. Respondents who rated the credit management policies as strong formed the largest category with 36 representing 70.6%. This is seconded by the opinion of those who considered it moderate as with 10 representing 19.6%. However variable 'Poor' accounted for the least with 5 denoting 9.8%. This implies that the banks' credit management policies are averagely good and strong. The fact that the majority of the respondents viewed character appraisal of the client of the banks as fair, indicates effective credit management.

Table 11. Rate of Credit Management under Collateral Appraisal of Clients

| Variable | Frequency | Percentage |
|---|---|---|
| Poor | 1 | 2.0 |
| Moderate | 8 | 15.7 |
| Strong | 42 | 82.4 |
| Total | 51 | 100.0 |

*Data Source: Survey Work 2020*

Table 11 also reveals the rate of credit management of microfinance banks under collateral of clients. Those that have rated the credit management policy as strong recorded the highest scores of 42 denoting 82.4%. It was succeeded by those who rated the policy as moderate with 8 connoting 15.7%. However, only 1 respondent representing 2.0% viewed the policy as poor. This certifies that the banks are strongly good in the collateral appraisal of clients. The fact that they are good in the collateral appraisal of clients can support the banks to manage their credit facility very well and healthy.

Table 12. Rate of Credit Management under Capital Appraisal of Clients

| Variable | Frequency | Percentage |
|---|---|---|
| Poor | 1 | 2.0 |
| Moderate | 10 | 19.6 |
| Strong | 40 | 78.4 |
| Total | 51 | 100.0 |

*Data Source: Survey Work 2020*

Table 12 above depicts the respondents' rate of credit management of microfinance banks under capital base appraisal of clients. Variable 'strong' accounted for the highest with 40 respondents representing 78.4%. It was seconded by variable 'moderate' with 10 people connoting 19.6%. However variable 'poor' recorded the least score of 1 representing 2.0%. This means that the banks have a very strong policy in respect of capital base appraisal of clients. The fact that the majority of the respondents strongly revealed that capital appraisal of clients is strong is an indicator of effective credit management of microfinance banks.

Table 13. Rate of Credit Management under Capacity Appraisal of Clients

| Variable | Frequency | Percentage |
|---|---|---|
| Poor | 2 | 3.9 |
| Moderate | 7 | 13.7 |
| Strong | 42 | 82.4 |
| Total | 51 | 100.0 |

*Data Source: Survey Work 2020*

Table 13 describes the rate of credit management of microfinance banks under capacity appraisal of clients. Respondents who considered capacity appraisal of clients as strong, and a factor that is helping the banks in credit management very well marked as the overall with 42 representing 82.4%. Only 7 individuals constituting 13.7% viewed it as moderate. However variable 'poor' recorded 2 scores representing 3.9%. This portrays that the majority of the respondents believed that capacity appraisal of clients is excellent. A healthy capacity appraisal of the clients is an effective tool for banks in credit management.

Table 14. Rate of Credit Management under Condition Appraisal of Clients

| Variable | Frequency | Percentage |
|---|---|---|
| Poor | 0 | 00.0 |
| Moderate | 17 | 33.3 |
| Strong | 34 | 66.7 |
| Total | 51 | 100.0 |

*Data Source: Survey Work 2020*

Table 14 reveals how respondents rated the credit management policy of their respective banks under condition appraisal of the client. Respondents who rated the credit management policies as strong formed the largest with 34 representing 66.7%. They were followed by the opinion of those who considered it as moderate with 17 representing 33.3%. However variable 'Poor' scored zero, meaning that the condition appraisal of clients of the banks is strong and rich. Strong conditions of appraisal of clients by microfinance banks are the yardstick of very efficient credit management policies.

### 4.1 Testing of Hypothesis

As earlier stated that the established hypotheses with regard to this study will be tested using multiple regressions. However, prior to the test of each hypothesis, Pearson's correlation of both one and two-tailed significance was employed to explain the association between the regresses and the regressors as well as the association between the regressors themselves. Similarly, the Kolmogorov – Smirnov normality test was used to test whether the sample distribution is significantly different from a normal distribution. The tests were presented below;

Table 15. Pearson's Correlation Matrix

| | The efficiency of Bank before Client Appraisal | Bank Client Appraisal on Character | Bank Client Appraisal on Collateral | Bank Client Appraisal on Capital | Bank Client Appraisal on Capacity | Bank Client Appraisal on Condition |
|---|---|---|---|---|---|---|
| Efficiency of Bank before Client Appraisal | 1 | -.348* | -.100 | -.355* | -.098 | .557** |
| | | .012 | .487 | .011 | .496 | .000 |
| | 51 | 51 | 51 | 51 | 51 | 51 |
| Bank Client Appraisal on Character | -.348* | 1 | .408** | .305* | .101 | -.105 |
| | .012 | | .003 | .030 | .481 | .463 |
| | 51 | 51 | 51 | 51 | 51 | 51 |
| Bank Client Appraisal on Collateral | -.100 | .408** | 1 | .042 | .253 | -.125 |
| | | .487 | .003 | | .772 | .074 | .382 |
| | | 51 | 51 | 51 | 51 | 51 | 51 |
| Bank Client Appraisal on Capital | -.355* | .305* | .042 | 1 | -.066 | -.290* |
| | .011 | .030 | .772 | | .645 | .039 |
| | 51 | 51 | 51 | 51 | 51 | 51 |
| Bank Client Appraisal on Capacity | -.098 | .101 | .253 | -.066 | 1 | -.307* |
| | .496 | .481 | .074 | .645 | | .029 |
| | 51 | 51 | 51 | 51 | 51 | 51 |
| Bank Client Appraisal on Condition | .557** | -.105 | -.125 | -.290* | -.307* | 1 |
| | .000 | .463 | .382 | .039 | .029 | |
| | 51 | 51 | 51 | 51 | 51 | 51 |

*. Correlation is significant at the 0.05 level (2-tailed).

**. Correlation is significant at the 0.01 level (2-tailed).

Table 15. above shows that all the independent variables (Appraisal of clients' base on character, collateral, capital, and capacity) are negatively associated with the dependent variables (Efficiency and productivity of banks before clients' appraisal). Except for appraisal of clients' base on condition that is positively associated with the dependent variable (Efficiency and productivity of banks before clients' appraisal). However, appraisal of clients' base on character and capital are jointly significantly related with the dependent variable at a 1% level of significance indicating a strong, negative relationship. Meanwhile, appraisal of clients' base on conditions is strongly, positively, and significantly related to the dependent variable at a 2% level of significance. While for appraisal of clients' base on collateral and capacity with dependent variable were insignificantly related. Among the five exogenous variables, the relationship was very strong as expected except for only two independent variables that were insignificantly related. Although some of the independent variables are negatively related, some were positively related.

Table 16. One-Sample Kolmogorov-Smirnov Test

a. Test distribution is Normal.

| | | Efficiency of Bank before Client Appraisal | Bank Client Appraisal on Character | Bank Client Appraisal on Collateral | Bank Client Appraisal on Capital | Bank Client Appraisal on Capacity | Bank Client Appraisal on Condition |
|---|---|---|---|---|---|---|---|
| N | | 51 | 51 | 51 | 51 | 51 | 51 |
| Normal Parametersa | Mean | 1.8431 | 2.6078 | 2.8039 | 2.7451 | 2.7843 | 2.6667 |
| | Std. Deviation | .70349 | .66569 | .44809 | .48345 | .50254 | .47610 |
| Most Extreme Differences | Absolute | .255 | .428 | .493 | .466 | .490 | .425 |
| | Positive | .235 | .278 | .331 | .299 | .334 | .253 |
| | Negative | -.255 | -.428 | -.493 | -.466 | -.490 | -.425 |
| Kolmogorov-Smirnov Z | | 1.820 | 3.056 | 3.518 | 3.326 | 3.497 | 3.033 |
| Asymp. Sig. (2-tailed) | | .003 | .000 | .000 | .000 | .000 | .000 |

Table 16 reveals the Kolmogorov - Smirnov normality test which tests whether the sample distribution is significantly different from a normal distribution. Thus, a significant p-value indicates a non – normal distribution of data. Therefore, going by the above table, it becomes apparent that the data are not normally distributed as the p-value of the Kolmogorov – Smirnov (0.003) and (0.000) are less than 0.05 for the data. However, the test distribution table suggests that the test distribution is normal.

Hypothesis: Client Appraisal has no significant effect on the Efficiency and Productivity of Microfinance Banks.

Table 17. Model Summary 1

| Model | R | R Square | Adjusted R Square | Std. Error of the Est. |
|---|---|---|---|---|
| 1 | 0.819 | 0.671 | 0.561 | 0.45017 |

*Data Source: SPSS Version 17 Computation*

Table 17 indicates that 'R' reveals the relationship between independent variables (Appraisal of client base on character, collateral, capital, capacity, and condition) and dependent variable (Efficiency and productivity of banks). The rule states that, the closer the figure to 1 the stronger the relationship and vice versa. Therefore with regard to this model, the relationship between the independent and dependent variable is strong with 0.81. This implies that the model has a strong relationship and goodness fit; meaning that appraisal of client character, collateral, capital, capacity, and condition before giving a loan has a positive effect on the efficiency and productivity of banks. Equally, the model reveals an R2 of 0.671, meaning that about a 67% increase in the efficiency and productivity of microfinance banks is accounted for by the variables in the model while the remaining 33% is accounted for by other factors not captured by the model. The robustness and goodness of fit of the model are further confirmed by an adjusted R2 of 0.561, which implies that 56% of the variation in the dependent variable is accounted for by the regresses.

Table 18. Model Summary 2

| Model | Sum of Squares | DF | Mean Square | F | Sig. |
|---|---|---|---|---|---|
| Regression | 10.769 | 5 | 2.154 | 6.934 | 0.000 |
| Residual | 13.976 | 45 | 0.311 | | |
| Total | 24.745 | 50 | | | |

*Data Source: SPSS Version 17 Computation*

Table 18 uses F – Statistics to tests for the overall significance of the hypothesis and regressor of the study. The regressor is significant at both 1% and 5% levels of significance. The rule says if the calculated F value is greater than the tabulated F value rejects the null hypothesis. Therefore, with regard to this model the F calculated is 6.934 while F tabulated or significant value is 0.000. Since F calculated is greater than F tabulated, so we reject the null hypothesis of no significance and conclude that client appraisal has a significant effect on the efficiency and productivity of microfinance banks.

Table 19: Model Summary 3

| Model | Coefficient | T – Statistics | Sig. Value |
|---|---|---|---|
| 1 (Constant) | 4.634 | 2.684 | 0.017 |
| Credit Mgt. under Character Appraisal | -0.399 | -1.458 | 0.166 |
| Credit Mgt. under Collateral Appraisal | -0.037 | -0,164 | 0.001 |
| Credit Mgt. under Capital Appraisal | 0.732 | 3.992 | 0.001 |
| Credit Mgt. under Capacity Appraisal | 0.150 | 0.713 | 0.212 |
| Credit Mgt. under Condition Appraisal | 0.878 | 4.907 | 0.103 |

*Data Source: SPSS Version 17 Computation*

Table 19 presents the effects of the independent variable (Appraisal of client base on character, collateral, capital, capacity, and condition) and dependent variable (Efficiency and productivity of banks). But the magnitude of the effects is of varying degree. Credit management under condition appraisal and nature of clients has a very strong positive effect given a coefficient of 0.878. This indicates that appraising the client base on conditions and nature of clients will bring about a change in efficiency and productivity of bank credit management by up to 88%. Similarly, credit management with a capital appraisal of a client has a strong positive effect given a coefficient of 0.732. This implies that appraising the client base on capital will bring about a change in efficiency and productivity of bank credit management by up to 73%. Meanwhile, credit management with regard to the capacity of clients has a relatively positive change of 0.150. This means that appraising the client base on capacity will bring about a marginal change in efficiency and productivity of bank credit management by up to 15%. However, based on prior knowledge of the above table, credit management with client appraisal on collateral and character showed negative effects. Although the negative effects of such variables were revealed to be negligible and low, given coefficients of -0.02% and -0.39%. The multiple regression results have also revealed that the Conditions appraisal of the clients is the most statistically significant of the repressors' with a T – statistics of 4.907 at a 1% level of significance. Capital appraisal of the clients by the banks is also statistically significant with a T – statistics of 3.992 at a 1% level of significance. In the same vein capacity appraisal of the clients by the microfinance banks is also found to be statistically significant with a T – statistics of 0.713 at a 1% level of significance. Two variables (collateral and character) were found to be statistically insignificant at a 1% level of significance given T – statistics of -0.164 and -1.458. However, the general outcome of the multiple regression models (results) showed a positive effect between the independent and dependent variables.

Table 20. Pearson's Correlation Matrix

|  | Profitability of Bank before Credit Risk Control | Bank Credit Risk Control Measures on Loan Product Design | Bank Credit Risk Control Measures on Credit Committee | Bank Credit Risk Control Measures Delinquency Management |
|---|---|---|---|---|
| Profitability of Bank before Credit Risk Control | 1 | .282* | .349* | .196 |
|  |  | .045 | .012 | .167 |
|  | 51 | 51 | 51 | 51 |
| Bank Credit Risk Control Measures on Loan Product Design | .282* | 1 | .319* | .758** |
|  | .045 |  | .022 | .000 |
|  | 51 | 51 | 51 | 51 |
| Bank Credit Risk Control Measures on Credit Committee | .349* | .319* | 1 | .568** |
|  | .012 | .022 |  | .000 |
|  | 51 | 51 | 51 | 51 |
| Bank Credit Risk Control Measures Delinquency Management | .196 | .758** | .568** | 1 |
|  | .167 | .000 | .000 |  |
|  | 51 | 51 | 51 | 51 |

*. Correlation is significant at the 0.05 level (2-tailed).

**. Correlation is significant at the 0.01 level (2-tailed).

Table 20 above reveals that all the independent variables ((Credit risk control measures under loan product design, credit committee, and delinquency management) are positively related to the dependent variable (Profitability of microfinance banks). However, credit risk control measures under loan product design and credit committee are jointly significantly related with the dependent variable at a 1% level of significance indicating a strong, positive relationship. However, credit risk control measures under loan delinquency management with the dependent variable are insignificantly related. Among the three exogenous variables, the relationship was very strong as expected except for one independent variable that is insignificantly related. Meanwhile, all the independent variables are positively related.

Table 21 depicts the Kolmogorov - Smirnov normality test which tests whether the sample distribution is significantly different from a normal distribution. Thus, a significant p-value indicates a non – normal distribution of data. Therefore, going by the above table, it becomes apparent that the data are not normally distributed giving a p-value of the Kolmogorov – Smirnov (0.000) and is less than 0.05 for the data. Although the test distribution table indicates that the test distribution is normal.

Table 21. One-Sample Kolmogorov-Smirnov Test

|  |  | Profitability of Bank before Credit Risk Control | Bank Credit Risk Control Measures Adopted 1 | Bank Credit Risk Control Measures Adopted 2 | Bank Credit Risk Control Measures Adopted 3 |
|---|---|---|---|---|---|
| N |  | 51 | 51 | 51 | 51 |
| Normal Parametersa | Mean | 1.5882 | 2.8235 | 2.7451 | 2.7843 |
|  | Std. Deviation | .60585 | .38501 | .44014 | .41539 |
| Most Extreme Differences | Absolute | .305 | .500 | .464 | .483 |
|  | Positive | .305 | .323 | .281 | .302 |
|  | Negative | -.281 | -.500 | -.464 | -.483 |
| Kolmogorov-Smirnov Z |  | 2.177 | 3.572 | 3.313 | 3.446 |
| Asymp. Sig. (2-tailed) |  | .000 | .000 | .000 | .000 |

a. Test distribution is Normal.

## V. CONCLUSION

The survivals of any financial institution depend on the available and sound liquidity of such an institution. Sound and healthy cash liquidity cannot be achieved without a good credit management policy. A good credit management policy guarantees healthy, sustainable, and profitable financial institutions. It also minimizes and reduces the amount of capital tied up with debtors and poor financial performance. Although several studies were conducted on the effect of credit management on the financial performance of financial institutions in a different dimension, however, this study focuses on microfinance banks in Adamawa state. With regard to this, the multiple regression models employed to assess the effect of client appraisal on the efficiency of microfinance banks revealed that client appraisal has a positive effect. This implies that client appraisal helps the microfinance banks to be efficient and productive in credit management. However, this finding is contrary to the findings of Gambo (2012) that revealed that client appraisal with regard to capability, gender and geographical location does not affect the credit management of microfinance banks. This study recommends that microfinance banks should maintain the appraisal of their clients' base on their capacity, capital base, collateral, character, condition etc. before giving the loans so as to recover the loans easily, thereby leading to the effectiveness and efficiency of banks. This is because client appraisal proved to be a very efficient yardstick in the credit management of banks and their financial performance. In addition, Microfinance banks should be reviewing and re-strategizing their client appraisal policies at the end of every financial year so as to improve their efficiency. This is because client appraisal proved to be an effective tool of credit management in improving the financial performance of the banks.

*\*Corresponding Author: Esther Yusuf Enoch[1]*
[1]*(Department of Banking and Finance, Federal Polytechnic Mubi, Adamawa State, Nigeria*